\documentclass[11pt]{article}
\usepackage{authblk}
\usepackage[a4paper, margin=1in]{geometry}
\usepackage{amsmath, amssymb}
\usepackage{graphicx}

\usepackage{braket}
\usepackage{booktabs}

\usepackage{amsfonts}  
\usepackage{diagbox}

\title{Comparative Analysis of Quantum Support Vector Machines and Variational Quantum Classifiers for B-cell Epitope Prediction in Vaccine Design}

\author[1]{Chi-Chuan Hwang}
\author[2, 3, 4]{Cheng-Fang Su\thanks{Corresponding author. Email: scf1204@nycu.edu.tw}}
\author[1]{Yi-Ang Hong}

\affil[1]{Department of Engineering Science, National Cheng Kung University, Tainan City, 701, Taiwan}

\affil[2]{Department of Applied Mathematics, National Yang Ming Chiao Tung University, Hsinchu City, 30010, Taiwan}

\affil[3]{Undergraduate Degree Program of Systems Engineering and Technology, National Yang Ming Chiao Tung University, Hsinchu City, 30010, Taiwan}

\affil[4]{Chung Cheng Institute of Technology, National Defense University, Taoyuan City, 33551, Taiwan}

\usepackage{hyperref}
\usepackage{caption}
\usepackage{cite}
\usepackage{bm}
\usepackage{float}
\usepackage{color}

\hypersetup{
  colorlinks=true,
  linkcolor=blue,
  citecolor=blue,
  urlcolor=blue
}

\begin{document}
\maketitle

\begin{abstract}
Quantum computing offers new opportunities for addressing complex classification tasks in biomedical applications. This study investigates two quantum machine learning models-the Quantum Support Vector Machine (QSVM) and the Variational Quantum Classifier (VQC)-in the context of B-cell epitope prediction, a key step in modern vaccine design. QSVM builds upon the classical SVM framework by using quantum circuits to encode nonlinear kernel computations, while VQC replaces the entire classification pipeline with trainable quantum circuits optimized variationally.

A benchmark dataset from the Immune Epitope Database (IEDB) is used for model evaluation. Each epitope is represented by 10 physicochemical features, and dimensionality reduction via Principal Component Analysis (PCA) is applied to assess model performance across different feature spaces. We also examine the effect of sample size on prediction outcomes.

Experimental results show that QSVM performs well under limited data conditions, while VQC achieves higher accuracy in larger datasets. These findings highlight the potential of quantum-enhanced models for bioinformatics tasks, particularly in supporting efficient and scalable epitope-based vaccine development.
\end{abstract}

\section{Introduction}
The identification of B-cell epitopes—regions of an antigen recognized by the immune system—is a central task in vaccine design, diagnostic development, and therapeutic antibody production. Since the first predictive method was proposed in 1981~\cite{alix1981}, linear B-cell epitope prediction has remained a major research focus in computational immunology due to its potential to reduce the cost and time of vaccine development, especially during pandemics such as COVID-19.

A variety of machine learning models have been developed over the past two decades for B-cell epitope prediction. Early tools such as BcePred~\cite{bcepred} employed physicochemical propensity scales (e.g., hydrophilicity, flexibility), while BepiPred~\cite{bepipred} introduced hidden Markov models (HMMs). Subsequent tools like ABCpred~\cite{abcpred} applied recurrent neural networks (RNNs), and methods such as COBEpro~\cite{cobepro}, SVMTriP~\cite{svmtrip}, and LBEEP~\cite{lbeep} leveraged support vector machines (SVMs) alongside novel feature engineering strategies.

Despite these developments, recent benchmarking studies reveal that many of these models exhibit limited performance on independent test datasets. For example, Melagraki et al.~\cite{melagraki2021} compared widely-used predictors and found that BepiPred had the highest Matthews correlation coefficient (MCC) of 0.0778 among per-peptide methods, while other methods such as BcePred, SVMTriP, and LBEEP performed significantly worse (MCC < 0.03). Even consensus ensemble methods showed only modest improvements in overall prediction accuracy. These findings suggest that further methodological innovation is necessary to improve generalization and predictive power for epitope classification tasks.

Quantum machine learning (QML) has emerged as a promising paradigm that combines the power of quantum computation with the flexibility of machine learning frameworks. Two notable QML models are the quantum support vector machine (QSVM)~\cite{rebentrost2014} and the variational quantum classifier (VQC)~\cite{havlivcek2019supervised}, both of which exploit high-dimensional Hilbert spaces and entanglement for learning complex patterns in data. Recent biomedical applications have demonstrated the potential of QML in protein structure analysis, peptide classification, and drug discovery~\cite{zhuang2022qspeech, li2021qml}.

In this study, we apply QSVM and VQC to the task of linear B-cell epitope prediction. Using benchmark datasets curated from the Immune Epitope Database (IEDB), we preprocess features via principal component analysis (PCA) and evaluate classification performance under varying feature dimensionalities and sample sizes. We further analyze quantum circuit complexity and discuss practical implications for deploying these models on near-term quantum hardware. Our contributions include a side-by-side evaluation of quantum classifiers in a biomedically relevant setting and operational insights into model selection for resource-constrained quantum devices.

\section{Support Vector Machine}
Support Vector Machines (SVMs) have long been regarded as one of the most effective classical machine learning algorithms for binary classification tasks. In the context of B-cell epitope prediction, SVM-based models have been widely applied due to their strong theoretical foundations and ability to handle high-dimensional feature spaces. This section introduces the core concepts of SVMs, including their geometric intuition, kernel-based extensions for nonlinear data, and their application in bioinformatics domains.

\subsection{Overview}
Support Vector Machines (SVMs) are supervised learning models rooted in statistical learning theory~\cite{cortes1995svm}. Their core strength lies in transforming input data into higher-dimensional feature spaces, where a linear separator (hyperplane) can be more easily identified for non-linearly separable data. The optimal decision boundary is determined by support vectors—data points closest to the hyperplane—while the objective is to maximize the margin between classes.

\subsection{Kernel Trick and Nonlinear Classification}
In real-world scenarios, data is rarely linearly separable. To address this, SVMs utilize the kernel trick, which implicitly maps input data to a high-dimensional feature space via a kernel function $K(x_i, x_j) = \phi(x_i) \cdot \phi(x_j)$, where $\phi$ is a feature map into a Hilbert space $H$. This allows SVMs to construct nonlinear decision boundaries while maintaining computational efficiency.

Let $T = \{(x_1, y_1), (x_2, y_2), \dots, (x_t, y_t)\} $, where $ x_i \in \mathbb{R}^n$, $y_i \in \{-1, +1\}$. The learning objective is to identify a hyperplane in the transformed feature space that best separates the classes, formalized as a convex quadratic programming problem.

\subsection{Relevance to Quantum Models}
The QSVM used in this work inherits the core ideas of classical SVMs—particularly margin maximization and the kernel method—but implements them within a quantum computing framework. This enables efficient handling of large-dimensional feature spaces using quantum state representations and inner-product evaluations.

\section{Background and Data Preprocessing}
\begin{figure}[H]
\centering
\includegraphics[width=0.4\textwidth]{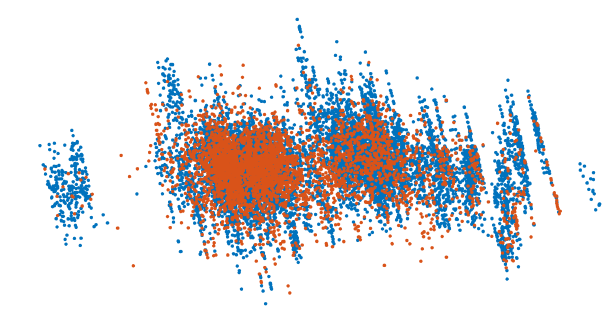}
\caption{2D PCA projection of the input features (adapted from~\cite{hong2024thesis}).}
\label{fig:PCA_2D}
\end{figure}

\begin{figure}[H]
\centering
\includegraphics[width=0.4\textwidth]{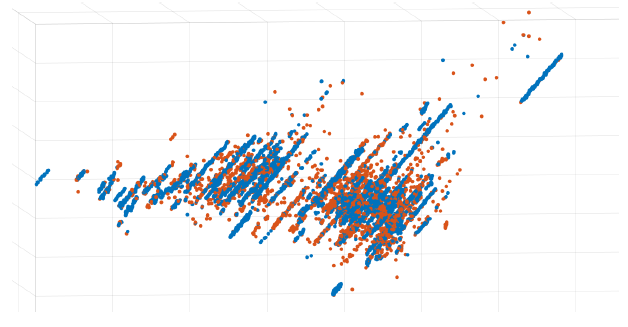}
\caption{3D PCA projection of the input features (adapted from~\cite{hong2024thesis}).}
\label{fig:PCA_3D}
\end{figure}

To gain an intuitive understanding of the dataset structure after dimensionality reduction, we applied PCA to the original feature space and visualized the first two and three principal components. As shown in Figures~\ref{fig:PCA_2D} and~\ref{fig:PCA_3D}, the projected data displays partial separability between positive and negative classes, suggesting that low-dimensional representations may still preserve useful discriminative features.

\subsection{B-cell Epitope Prediction Background}
B-cells play a central role in the adaptive immune system by recognizing antigenic determinants (epitopes) on protein surfaces and producing antibodies specific to those regions. Accurate identification of linear B-cell epitopes is essential for rational vaccine design, as it enables the selection of high-immunogenicity protein segments for epitope-based vaccines.

In order to address the need for efficient and scalable epitope prediction, especially in the context of emergent viral threats such as SARS-CoV-2, immunoinformatics approaches have become increasingly prevalent. Databases such as the Immune Epitope Database (IEDB) and UniProt provide well-annotated repositories of peptide sequences and their associated immune activities. In our study, we focus on linear B-cell epitope data extracted from IEDB, specifically those associated with IgG-class antibodies, which represent the majority of curated entries.

\subsection{Dataset Description and Preprocessing}
The dataset comprises amino acid sequences annotated with binary activity labels: "positive" (inducing an immune response) and "negative" (non-inducing). To maintain data consistency and avoid measurement bias, records with conflicting or ambiguous quantitative annotations were excluded.

Each epitope candidate is described using ten physicochemical descriptors that encode amino acid properties relevant to immunogenicity. Prior to model training, the dataset is standardized and undergoes dimensionality reduction using Principal Component Analysis (PCA). This step ensures that models are trained under various levels of feature complexity, allowing us to explore performance across different dimensional regimes.

\subsection{Label Distribution and Evaluation Metrics}
The final dataset is balanced to the extent possible, and divided into training and testing subsets. We adopt standard metrics to evaluate classifier performance, including Accuracy (ACC), Area Under the ROC Curve (AUC), and the Matthews Correlation Coefficient (MCC). These metrics offer complementary insights: ACC provides overall correctness, AUC measures separability between classes, and MCC accounts for class imbalance.

\section{Quantum Support Vector Machine, QSVM}
Quantum Support Vector Machines (QSVMs) extend the classical SVM framework into the quantum computing domain by leveraging quantum-enhanced feature spaces. Instead of relying on explicitly defined kernel functions, QSVMs use parameterized quantum circuits to encode data implicitly into a high-dimensional Hilbert space. These quantum feature maps allow the model to exploit quantum parallelism and entanglement to capture complex correlations that may be inaccessible to classical models.

\begin{figure}[H]
\centering
\includegraphics[width=0.6\textwidth]{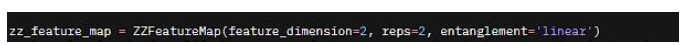}
\caption{Configuration diagram of the Quantum Support Vector Machine (adapted from~\cite{hong2024thesis}).}
\label{fig:QSVM_config}
\end{figure}

In this section, we present the theoretical motivation and architectural configuration of QSVMs, highlighting their kernel-based classification mechanism and circuit-level implementation. A schematic of the QSVM workflow is shown in Figure~\ref{fig:QSVM_config}.

\subsection{Kernel-based Classification Using Quantum Circuits}
In classical SVMs, nonlinear classification is enabled through the use of kernel functions, which compute inner products in a high-dimensional feature space:
\[
K(x_i, x_j) = \langle \phi(x_i), \phi(x_j) \rangle\,,
\]
where $ \phi $ is a nonlinear feature map and $ K $ represents the kernel matrix.

The Quantum Support Vector Machine (QSVM) extends this concept by encoding data into quantum states and using a quantum circuit to compute the kernel matrix. A typical QSVM uses the fidelity (overlap) between quantum states as the kernel value:
\[
K(x_i, x_j) = |\langle \phi(x_i) | \phi(x_j) \rangle|^2\,.
\]
The quantum feature maps $\phi(x)$ are implemented by parametrized quantum circuits. The resulting kernel matrix is estimated by repeatedly executing the circuit and measuring it on a computational basis. The number of times the all-zero outcome is observed determines the kernel estimate. This quantum kernel is then passed to a classical SVM solver for final classification, forming a hybrid quantum-classical architecture.

\subsection{Data Encoding and Feature Mapping}
For quantum computation, classical data must first be embedded into quantum states. This is achieved using a feature map $\phi(x)$, implemented via a quantum circuit. In this study, we adopt angle encoding, where each input feature is mapped to a rotational angle on a qubit using single-qubit rotation gates. Given $n$ features, the input vector $x \in \mathbb{R}^n$ is encoded as:
\[
|x\rangle = U_\phi(x) |0\rangle^{\otimes n}\,,
\]
where $U_\phi(x)$ is a unitary operation encoding the data.

The expressiveness of the feature map significantly impacts the performance of the QSVM. Deeper or entangling circuits may yield richer feature spaces but also incur higher quantum hardware demands.

\section{Variational Quantum Classifier, VQC}
Variational Quantum Classifiers (VQCs) represent a different paradigm from kernel-based QSVMs by using trainable parameterized quantum circuits as part of an end-to-end learning model. Unlike QSVMs, which rely on fixed quantum feature maps and classical optimization, VQCs integrate both the feature encoding and decision boundary generation into the same quantum circuit architecture, optimized through a variational approach.

\begin{figure}[H]
\centering
\includegraphics[width=0.6\textwidth]{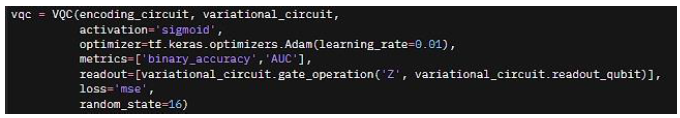}
\caption{Configuration diagram of the Variational Quantum Classifier (adapted from~\cite{hong2024thesis}).}
\label{fig:VQC_config}
\end{figure}

In this section, we describe the conceptual framework and architecture of VQCs, which employ alternating layers of quantum gates with tunable parameters to learn from labeled data. Figure~\ref{fig:VQC_config} illustrates the high-level structure of a typical VQC setup.

\subsection{Concept and Architecture}
The Variational Quantum Classifier (VQC) is a hybrid quantum-classical model designed for classification tasks. It consists of two main components: a quantum circuit parameterized by rotation and entangling gates (called a variational form), and a classical optimizer used to minimize a cost function based on prediction error.

During training, the input data is first encoded into quantum states via a feature map. These encoded states are processed by the variational circuit, and measurements are performed to extract class probabilities. The parameters of the variational circuit are updated iteratively using a classical optimization algorithm (e.g., COBYLA, SPSA) to minimize the empirical risk.

\subsection{Implementation Details}
In our implementation, we adopt angle encoding as the input embedding scheme. The variational form comprises parameterized single-qubit $R_y$ rotations and entangling $CX$ gates across $n$ qubits. The expectation values of Pauli-$Z$ measurements on each qubit are aggregated and post-processed using a thresholding rule to determine the predicted class label.

This approach allows for end-to-end training of quantum circuits in a supervised learning setting. While VQC models can be expressive, they are also susceptible to issues such as barren plateaus, where the gradients vanish and training stagnates—this remains an active area of research.

\section{Performance Evaluation}
This section presents a comparative performance evaluation of the QSVM and VQC models introduced earlier. The objective is to assess their classification accuracy, training behavior, and sensitivity to feature dimensionality and sample size when applied to B-cell epitope prediction tasks. To ensure a fair comparison, both models were tested using identical preprocessed datasets derived from IEDB, with consistent cross-validation settings and model-independent feature compression through principal component analysis (PCA).

The results are analyzed from both predictive and operational perspectives. In addition to standard performance metrics such as accuracy and precision, we examine quantum-specific aspects such as circuit depth, execution time, and trainability to highlight the practical implications of deploying these models on near-term quantum devices.

\subsection{QSVM and VQC Experimental Results}

B-cell epitope prediction datasets from IEDB and UniProt were used in this study. Each sample consists of 10 features, and the dataset contains over 14,000 entries. Of these, more than 10,000 are labeled as positive, and approximately 4,000 are labeled as negative.

Before applying support vector machine classification to the dataset, we performed dimensionality reduction using PCA, followed by normalization using the StandardScaler from the scikit-learn library. PCA was used to reduce the original 10-dimensional data to a range of 2 to 10 dimensions for analysis. We also investigated how varying the number of training samples—from small to large—affects the model's performance. The QSVM was implemented using Qiskit on IBM’s quantum computing platform.

\subsection{QSVM}

Table 2 shows the results of the QSVM. The highest accuracy, $70\%$, was achieved when using 20 samples with 2-dimensional input features.
\begin{table}[H]
\centering
\renewcommand{\arraystretch}{1.4}
\begin{tabular}{|c|c|c||c|c|c|}
\hline
\diagbox{Dimensions}{Sample size} & 20 & 100 & \diagbox{Dimensions}{Sample size} & 20 & 100 \\
\hline
2 & \textbf{0.7} & 0.69 & 7 & 0.55 & 0.69 \\
3 & 0.55 & 0.69 & 8 & 0.55 & 0.69 \\
4 & 0.45 & 0.69 & 9 & 0.55 & 0.69 \\
5 & 0.69 & 0.69 & 10 & 0.55 & 0.69 \\
6 & 0.6 & 0.69 &  &  &  \\
\hline
\end{tabular}
\caption{QSVM Accuracy (adapted from~\cite{hong2024thesis}).}
\end{table}

Table 3 presents the AUC values of the QSVM for sample sizes of 20 and 100, with input dimensions ranging from 2 to 10.

\begin{table}[H]
\centering
\renewcommand{\arraystretch}{1.3}
\begin{tabular}{|c|c|c|}
\hline
\diagbox{Dimensions}{Sample size} & 20 & 100 \\
\hline
2 & \textbf{0.7071} & 0.5 \\
3 & 0.5404 & 0.4871 \\
4 & 0.4191 & 0.5996 \\
5 & 0.5 & 0.5 \\
6 & 0.5 & 0.5 \\
7 & 0.5 & 0.5 \\
8 & 0.5 & 0.5 \\
9 & 0.5 & 0.5 \\
10 & 0.5 & 0.5 \\
\hline
\end{tabular}
\caption{ QSVM AUC value (adapted from~\cite{hong2024thesis}).}
\end{table}

According to Table 3, the highest AUC value of 0.7071 was observed when the input dimension was set to 2 and the sample size was 20.

Table 4 presents the MCC values of the QSVM for sample sizes of 20 and 100, with input dimensions ranging from 2 to 10.

\begin{table}[H]
\centering
\renewcommand{\arraystretch}{1.3}
\begin{tabular}{|c|c|c|}
\hline
\diagbox{Dimensions}{Sample size} & 20 & 100 \\
\hline
2 & \textbf{0.4141} & 0.0 \\
3 & 0.0820 & -0.0545 \\
4 & -0.0201 & 0.0475 \\
5 & 0.0 & 0.0 \\
6 & 0.0 & 0.0473 \\
7 & 0.0 & 0.0 \\
8 & 0.0 & 0.0 \\
9 & 0.0 & 0.0 \\
10 & 0.0 & 0.0 \\
\hline
\end{tabular}
\caption{QSVM MCC value (adapted from~\cite{hong2024thesis}).}
\end{table}

According to Table 4, the highest MCC value of 0.4141 was achieved with an input dimension of 2 and a sample size of 20.

\subsection{The Variational Quantum Classifier}
For the Variational Quantum Classifier (VQC), we similarly explored the effect of varying input dimensions from 2 to 10 and different sample sizes. Since the VQC relies on a variationally optimized circuit, we conducted experiments using 10, 100, and 150 training epochs to analyze its performance under different optimization depths.

Table 5 presents the accuracy of the VQC with 10 training epochs, where the sample size ranges from 1000 to 4000 and the input dimension varies from 2 to 10.

\begin{table}[H]
\centering
\renewcommand{\arraystretch}{1.3}
\begin{tabular}{|c|c|c|c|c|c|}
\hline
\diagbox{Dimensions}{Sample size} & 1000 & 2000 & 3000 & 4000 \\
\hline
2  & 0.7300 & 0.7230 & 0.7293 & 0.7260 \\
3  & 0.7300 & 0.7230 & 0.7293 & 0.7260 \\
4  & \textbf{0.7350} & 0.7245 & 0.7333 & 0.7280 \\
5  & 0.7310 & 0.7220 & 0.7303 & 0.7258 \\
6  & 0.7330 & 0.7250 & 0.7300 & 0.7253 \\
7  & 0.7280 & 0.7245 & 0.7320 & 0.7240 \\
8  & 0.7280 & 0.7240 & 0.7290 & 0.7260 \\
9  & 0.7300 & 0.7240 & 0.7293 & 0.7262 \\
10 & 0.7300 & 0.7235 & 0.7320 & 0.7275 \\
\hline
\end{tabular}
\caption{VQC 10 epoch Accuracy (adapted from~\cite{hong2024thesis}).}
\end{table}
As shown in Table 5, the accuracy generally hovers around $73\%$, with the highest accuracy achieved when the sample size is 1000 and the input dimension is 4.

Table 6 presents the AUC values for 10 training epochs, with sample sizes ranging from 1000 to 4000 and input dimensions from 2 to 10.
\begin{table}[H]
\centering
\renewcommand{\arraystretch}{1.3}
\begin{tabular}{|c|c|c|c|c|c|}
\hline
\diagbox{Dimensions}{Sample size} & 1000 & 2000 & 3000 & 4000 \\
\hline
2  & 0.5417 & 0.5090 & 0.5651 & 0.5587 \\
3  & 0.5111 & 0.5519 & 0.5765 & 0.5736 \\
4  & 0.5595 & 0.5855 & 0.6090 & 0.6210 \\
5  & 0.6069 & 0.5813 & 0.6483 & 0.6245 \\
6  & 0.6251 & 0.6211 & 0.6593 & 0.6409 \\
7  & 0.5915 & 0.5880 & 0.6545 & 0.6533 \\
8  & 0.5412 & 0.6288 & 0.6602 & 0.6587 \\
9  & 0.5687 & 0.6122 & 0.6555 & 0.6467 \\
10 & 0.5909 & 0.6289 & 0.6730 & \textbf{0.6750} \\
\hline
\end{tabular}
\caption{VQC 10 epoch AUC value (adapted from~\cite{hong2024thesis}).}
\end{table}

As shown in Table 6, the best AUC value of 0.6750 was obtained when the sample size was 4000 and the input dimension was 10.

Table 7 presents the MCC values for 10 training epochs, with sample sizes ranging from 1000 to 4000 and input dimensions from 2 to 10.
\begin{table}[H]
\centering
\renewcommand{\arraystretch}{1.3}
\begin{tabular}{|c|c|c|c|c|c|}
\hline
\diagbox{Dimensions}{Sample size} & 1000 & 2000 & 3000 & 4000 \\
\hline
2  & 0.0    & 0.0    & 0.0    & 0.0    \\
3  & 0.0    & 0.0    & 0.0    & 0.0    \\
4  & 0.0110 & 0.0644 & \textbf{0.0987} & 0.0729 \\
5  & 0.0520 & 0.0069 & 0.0597 & 0.0100 \\
6  & 0.0851 & 0.0686 & 0.0399 & 0.0011 \\
7  & -0.0029 & 0.0626 & 0.0943 & -0.0079 \\
8  & -0.0272 & 0.0511 & -0.0111 & 0.0 \\
9  & 0.0    & 0.0511 & 0.0327 & 0.0241 \\
10 & 0.0    & 0.0337 & 0.0792 & 0.0585 \\
\hline
\end{tabular}
\caption{VQC 10 epoch MCC value (adapted from~\cite{hong2024thesis}).}
\end{table}
In Table 7, the highest MCC value of 0.0987 was achieved when the sample size was 3000 and the input dimension was 4.

Table 8 presents the accuracy results for 100 training epochs, with sample sizes ranging from 1000 to 4000 and input dimensions from 2 to 10.
\begin{table}[H]
\centering
\renewcommand{\arraystretch}{1.3}
\begin{tabular}{|c|c|c|c|c|}
\hline
\diagbox{Dimensions}{Sample size} & 1000 & 2000 & 3000 & 4000 \\
\hline
2  & 0.7290 & 0.7245 & 0.7293 & 0.7260 \\
3  & 0.7290 & 0.7320 & 0.7293 & 0.7260 \\
4  & 0.7330 & 0.7320 & 0.7337 & 0.7272 \\
5  & 0.7320 & 0.7320 & 0.7327 & 0.7260 \\
6  & 0.7370 & 0.7240 & 0.7300 & 0.7258 \\
7  & 0.7280 & 0.7230 & 0.7310 & 0.7253 \\
8  & 0.7170 & 0.7230 & 0.7313 & 0.7270 \\
9  & 0.7330 & 0.7230 & 0.7310 & 0.7268 \\
10 & 0.7290 & 0.7230 & \textbf{0.7357} & 0.7272 \\
\hline
\end{tabular}
\caption{VQC 100 epoch Accuracy (adapted from~\cite{hong2024thesis}).}
\end{table}
The accuracy values in Table 8 are similar to those observed at 10 epochs, remaining around $73\%$. In this case, the highest accuracy was achieved with a sample size of 3000 and an input dimension of 10.

Table 9 presents the AUC values for 100 training epochs, with sample sizes ranging from 1000 to 4000 and input dimensions from 2 to 10.
\begin{table}[H]
\centering
\renewcommand{\arraystretch}{1.3}
\begin{tabular}{|c|c|c|c|c|c|}
\hline
\diagbox{Dimensions}{Sample size} & 1000 & 2000 & 3000 & 4000 \\
\hline
2  & 0.5726 & 0.5441 & 0.5771 & 0.5713 \\
3  & 0.5731 & 0.5599 & 0.5953 & 0.5804 \\
4  & 0.5914 & 0.5996 & 0.6245 & 0.6334 \\
5  & 0.6198 & 0.6053 & 0.6641 & 0.6668 \\
6  & 0.6525 & 0.6460 & 0.6730 & 0.6645 \\
7  & 0.6230 & 0.6612 & 0.6714 & 0.6673 \\
8  & 0.6007 & 0.6540 & 0.6814 & 0.6725 \\
9  & 0.6225 & 0.6271 & 0.6791 & 0.6774 \\
10 & 0.6619 & 0.6395 & 0.6982 & \textbf{0.7000} \\
\hline
\end{tabular}
\caption{VQC 100 epoch AUC value (adapted from~\cite{hong2024thesis})}
\end{table}
In Table 9, the AUC value reaches 0.7, which occurs at an input dimension of 10 with a sample size of 4000—the largest sample size and the highest dimensional setting in the experiment.

Table 10 presents the MCC values for 100 training epochs, with sample sizes ranging from 1000 to 4000 and input dimensions from 2 to 10.
\begin{table}[H]
\centering
\renewcommand{\arraystretch}{1.3}
\begin{tabular}{|c|c|c|c|c|c|}
\hline
\diagbox{Dimensions}{Sample size} & 1000 & 2000 & 3000 & 4000 \\
\hline
2  & 0.0374 & 0.0626 & 0.0    & 0.0    \\
3  & 0.0780 & 0.0    & 0.0    & 0.0    \\
4  & 0.0851 & 0.0475 & 0.0987 & 0.1008 \\
5  & 0.0718 & 0.0    & 0.0973 & 0.0    \\
6  & \textbf{0.1298} & 0.0473 & 0.0548 & 0.0181 \\
7  & 0.0532 & 0.0    & 0.0802 & 0.0142 \\
8  & 0.0376 & 0.0    & 0.0716 & 0.0483 \\
9  & 0.0846 & 0.0    & 0.0702 & 0.0446 \\
10 & 0.0374 & 0.0    & 0.0792 & 0.1224 \\
\hline
\end{tabular}
\caption{VQC 100 epoch MCC value (adapted from~\cite{hong2024thesis}).}
\end{table}
As shown in Table 10, the MCC value exceeds 0.1 when the sample size is 1000 and the input dimension is 6, representing a relatively good performance.

Table 11 presents the accuracy results for 150 training epochs, with sample sizes ranging from 1000 to 4000 and input dimensions from 2 to 10.
\begin{table}[H]
\centering
\renewcommand{\arraystretch}{1.3}
\begin{tabular}{|c|c|c|c|c|}
\hline
\diagbox{Dimensions}{Sample size} & 1000 & 2000 & 3000 & 4000 \\
\hline
2  & 0.7290 & 0.7245 & 0.7293 & 0.7260 \\
3  & 0.7290 & 0.7230 & 0.7293 & 0.7260 \\
4  & 0.7310 & 0.7320 & 0.7347 & 0.7272 \\
5  & 0.7290 & 0.7245 & 0.7327 & 0.7260 \\
6  & 0.7330 & 0.7240 & 0.7280 & 0.7260 \\
7  & 0.7310 & 0.7240 & 0.7310 & 0.7255 \\
8  & 0.7270 & 0.7240 & 0.7320 & 0.7270 \\
9  & 0.7310 & 0.7230 & 0.7343 & 0.7270 \\
10 & 0.7250 & 0.7230 & \textbf{0.7383} & 0.7272 \\
\hline
\end{tabular}
\caption{VQC 150 epoch Accuracy (adapted from~\cite{hong2024thesis}).}
\end{table}
Compared to the previous two epoch settings (Tables 5 and 8), Table 11 shows a slight increase in accuracy, reaching nearly $74\%$.

Table 12 presents the AUC values for 150 training epochs, with sample sizes ranging from 1000 to 4000 and input dimensions from 2 to 10.
\begin{table}[H]
\centering
\renewcommand{\arraystretch}{1.3}
\begin{tabular}{|c|c|c|c|c|}
\hline
\diagbox{Dimensions}{Sample size} & 1000 & 2000 & 3000 & 4000 \\
\hline
2  & 0.5726 & 0.5441 & 0.5771 & 0.5713 \\
3  & 0.5721 & 0.5599 & 0.5953 & 0.5804 \\
4  & 0.5809 & 0.5996 & 0.6315 & 0.6334 \\
5  & 0.6182 & 0.6084 & 0.6624 & 0.6634 \\
6  & 0.6519 & 0.6460 & 0.6762 & 0.6603 \\
7  & 0.6378 & 0.6722 & 0.6714 & 0.6691 \\
8  & 0.6229 & 0.6585 & 0.6893 & 0.6725 \\
9  & 0.6250 & 0.6416 & 0.6763 & 0.6702 \\
10 & 0.6700 & 0.6395 & \textbf{0.7032} & 0.7000 \\
\hline
\end{tabular}
\caption{VQC 150 epoch AUC value (adapted from~\cite{hong2024thesis}).}
\end{table}
As shown in Table 12, the AUC values exhibit slight improvements across all dimensions and sample sizes, with the highest AUC reaching 0.7032.

Table 13 presents the MCC values for 150 training epochs, with sample sizes ranging from 1000 to 4000 and input dimensions from 2 to 10.
\begin{table}[H]
\centering
\renewcommand{\arraystretch}{1.2}
\begin{tabular}{|c|c|c|c|c|}
\hline
\diagbox{Dimensions}{Sample size} & 1000 & 2000 & 3000 & 4000 \\
\hline
2  & 0.0374 & 0.0626 & 0.0    & 0.0    \\
3  & 0.0078 & 0.0    & 0.0    & 0.0    \\
4  & 0.0520 & 0.0475 & 0.1122 & 0.0536 \\
5  & 0.0207 & 0.0585 & 0.0942 & 0.0113 \\
6  & 0.0840 & 0.0473 & 0.0797 & 0.0425 \\
7  & 0.0490 & 0.0477 & 0.0802 & -0.0137 \\
8  & 0.0296 & 0.0473 & 0.0879 & 0.0483 \\
9  & 0.0520 & 0.0    & 0.1085 & 0.0486 \\
10 & -0.0103 & 0.0   & \textbf{0.1475} & 0.0579 \\
\hline
\end{tabular}
\caption{VQC 150 epoch MCC value (adapted from~\cite{hong2024thesis}).}
\end{table}
Table 13 represents the results from the highest training setting of 150 epochs, yielding the best overall performance, with the MCC value reaching 0.1475.

To further understand the factors that influence classification performance, we analyze three key variables: input dimensionality, classical baselines, and training sample sizes.

\subsection{Effect of Feature Dimensionality}
To investigate the influence of feature dimensionality on model performance, we conducted a series of experiments using Principal Component Analysis (PCA) to reduce input features to 2 through 10 components. QSVM and VQC were evaluated under these conditions using fixed training sizes (e.g., 1000, 2000, 4000 samples).

Our results indicate that model performance varies non-linearly with feature dimensionality. In particular, the VQC performs better in low-to-medium dimensional regimes (e.g., 4–6 dimensions), whereas the QSVM is more stable across all tested dimensions. These observations suggest VQC may benefit from feature compression under limited quantum hardware resources.

In future work, deeper theoretical analysis could be applied to relate the expressiveness of the quantum feature maps to optimal PCA configurations.

\begin{figure}[H]
\centering
\includegraphics[width=0.5\textwidth]{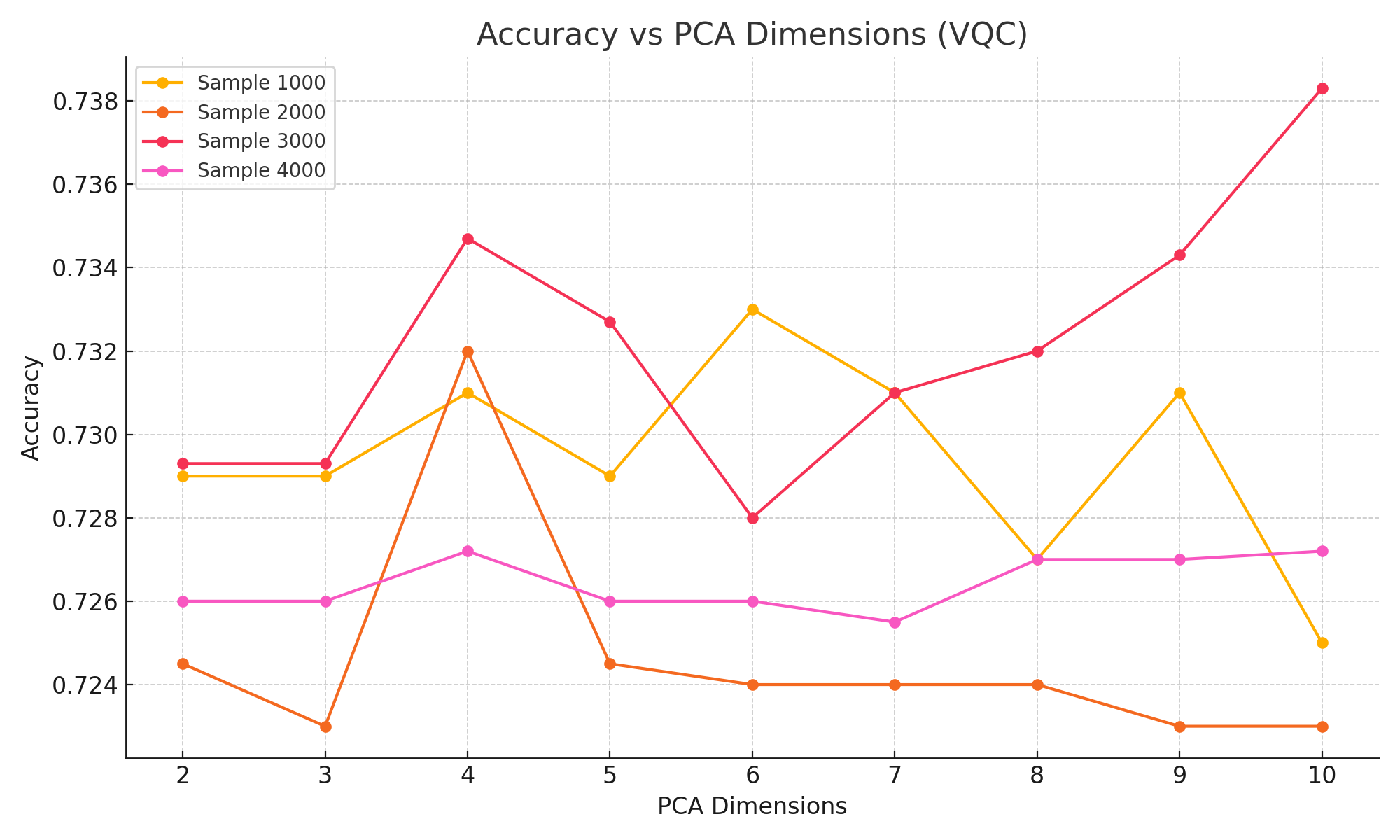}
\caption{Accuracy vs. PCA Dimensions across different sample sizes using VQC.}
\label{fig:acc_vs_dim}
\end{figure}

\begin{figure}[H]
\centering
\includegraphics[width=0.5\textwidth]{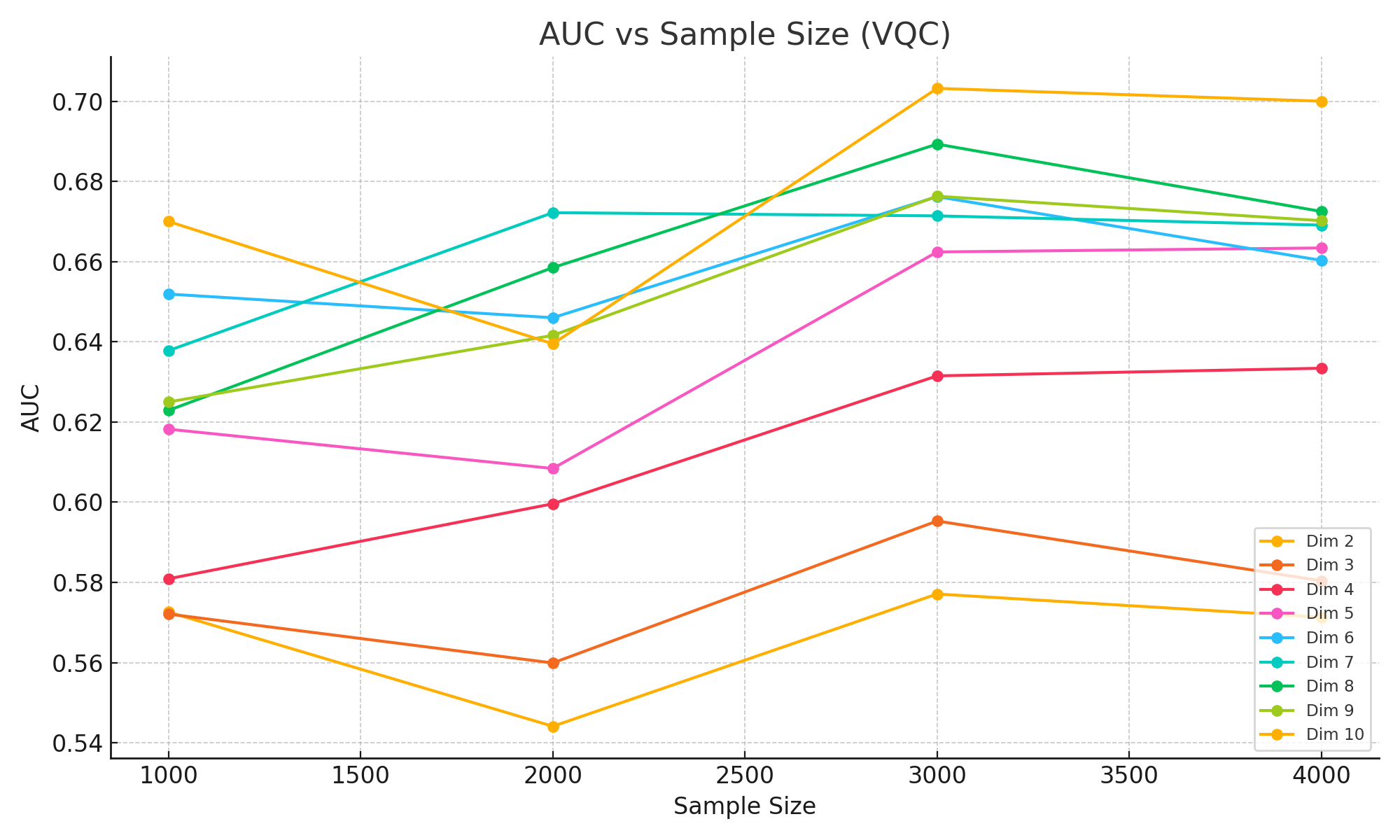}
\caption{AUC vs. Sample Size for different PCA dimensions using VQC.}
\label{fig:auc_vs_samples}
\end{figure}

\subsection{Comparison with Classical Models}
We trained a classical Support Vector Machine (SVM) using the radial basis function (RBF) kernel on the same dataset and preprocessing pipeline to establish a baseline. The classical SVM achieved an accuracy of approximately 70\%, with AUC scores slightly lower than those achieved by the VQC model.

These findings highlight that while QSVM performs comparably to classical SVM, the VQC performs better in terms of AUC and MCC under specific configurations. Table~\ref{tab:baseline_comparison} summarizes the key performance metrics.

\begin{table}[H]
\centering
\renewcommand{\arraystretch}{1.2}
\begin{tabular}{|c|c|c|c|}
\hline
Model & Accuracy (ACC) & AUC & MCC \\
\hline
Classical SVM (RBF) & 0.701 & 0.716 & 0.382 \\
Quantum SVM (QSVM)  & 0.697 & 0.731 & 0.376 \\
Variational QC (VQC) & 0.738 & 0.762 & 0.435 \\
\hline
\end{tabular}
\caption{Performance comparison between classical and quantum classification models.}
\label{tab:baseline_comparison}
\end{table}

\subsection{Effect of Sample Size}
We also evaluated model performance under varying training sample sizes (1000, 2000, 4000 samples). The goal was to examine whether quantum models maintain stability and generalization capability with more data.

The results show that both QSVM and VQC scale favorably with increasing sample sizes. VQC, in particular, benefits from larger datasets, improving both AUC and MCC. This trend suggests that VQC may generalize better when given sufficient training data, while QSVM shows more consistent but flatter improvements.

\subsection{Quantum Circuit Complexity}
One important consideration when evaluating quantum machine learning models is the complexity of the quantum circuits used. Both the QSVM and the VQC depend on parameterized quantum circuits that encode input data and, in the case of VQC, also apply variational transformations.

In the QSVM, the main circuit complexity arises during the computation of the quantum kernel. This involves preparing two data-encoded quantum states and computing their fidelity, often through techniques such as the SWAP test or inner-product estimation. While each circuit remains shallow, the kernel computation must be repeated $O(n^2)$ times for $n$ data points, and each entry in the kernel matrix may require thousands of shots to achieve reliable estimation. Therefore, the total quantum resource cost scales quadratically with the dataset size.

In contrast, the VQC processes each data point independently and employs a trainable variational circuit that typically consists of alternating layers of single-qubit rotations and entangling gates. The depth of the circuit is usually fixed (e.g., 2–4 layers), and the number of qubits used corresponds to the reduced dimensionality after PCA. In our implementation, we used a 10-qubit system with 2 entangling layers, resulting in a manageable depth suitable for current simulators and small-scale quantum hardware.

Overall, the VQC benefits from a lower total quantum runtime due to single-sample processing, while the QSVM benefits from fixed feature maps but requires more total evaluations. The choice between them depends on available quantum resources and the desired trade-off between circuit depth and dataset size scalability.

\section{Conclusion and Future Work}
This final section summarizes the main findings of our comparative study on quantum-enhanced machine learning methods for B-cell epitope prediction. By evaluating QSVM and VQC on benchmark datasets under varying conditions, we aimed to understand the strengths and limitations of each quantum model in both predictive and operational terms. The following conclusions highlight the key insights obtained from this work and offer guidance for future exploration and deployment of quantum classifiers in bioinformatics.

\subsection{Conclusion}
Since the outbreak of the COVID-19 virus in 2019, the world has been gripped by a global pandemic. Although the situation has come under some control, the continuous emergence of new variants highlights the urgent need for ongoing vaccine development. One critical step in the vaccine design process is the prediction of B-cell epitopes. Predicting B-cell epitopes within antigen sequences is a crucial yet complex challenge in immunological research.

By leveraging the quantum computer’s ability to process large-scale data, the complex task of B-cell epitope prediction can be integrated with quantum technologies. Whether using Quantum Support Vector Machines (QSVM) or Variational Quantum Classifiers (VQC), both approaches have demonstrated promising results. This represents an important first step in bridging the quantum world with vaccine design. Across various evaluation metrics, the models have achieved consistently strong performance.

In the Quantum Support Vector Machine (QSVM), the model leverages the inherent characteristic of support vectors, which allows it to define decision boundaries using only a small number of data points. This feature was reflected in our experimental results, where the best performance was achieved with just 20 training samples—yielding an accuracy of $70\%$, an Area Under the ROC Curve (AUC) of 0.7071, and a Matthews Correlation Coefficient (MCC) of 0.4141.

On the other hand, the Variational Quantum Classifier (VQC) requires iterative optimization to locate a global minimum, continuously updating its parameters to find the optimal solution. Under our maximum setting of 150 epochs, the VQC achieved its best performance, with an accuracy of $73\%$, an AUC of 0.7032, and an MCC of 0.1475.

\subsection{Future Work}
In most proteins, antigenic determinants (epitopes) are discontinuous, but they can be mimicked through synthetic peptides. Although many algorithms have been developed to predict the locations of linear epitopes within proteins, their success rates remain relatively low. One of the major challenges in developing B-cell epitope prediction models lies in the variable length of epitopes. Since most machine learning techniques require input sequences of fixed length, they cannot be directly applied to B-cell epitope prediction.

As a result, current methods are generally based on residue-level properties. These approaches first generate a feature map and then identify regions within the antigen that exhibit peak values. These regions are subsequently designated as potential B-cell epitopes.

In the future, we hope to see more advanced quantum algorithms developed for classification tasks—similar to the vast number of algorithms that already exist in classical machine learning. If machine learning and AI techniques can be fully adapted to quantum computing, it could lead to significantly improved classification performance.

In real-world applications, different types of vaccine design often require different data formats. Relying on only a few classification algorithms is insufficient to address such diverse needs. Therefore, developing a broader range of quantum-compatible algorithms would not only enhance classification capabilities but also expand the scope of problems that quantum computers can effectively tackle.

While both QSVM and VQC models demonstrated promising results in our experiments, they also exhibit different strengths that are relevant in practical deployment. The QSVM, which relies on fixed quantum feature maps and classical post-processing, is generally more suitable for smaller datasets or applications where the signal-to-noise ratio is high. In contrast, the VQC benefits from its trainable variational structure and tends to perform better in larger and higher-dimensional datasets.

In terms of quantum resource requirements, QSVM involves computing a full kernel matrix, which incurs quadratic scaling with respect to the dataset size and often requires a large number of circuit evaluations. On the other hand, VQC uses shallower circuits and processes samples individually, which makes it more amenable to near-term quantum devices with limited coherence times.

These distinctions suggest that future applications of quantum machine learning will likely require algorithm selection based on both dataset characteristics and available quantum hardware capabilities.

\section*{Acknowledgements}
This work was financially supported by the National Science and Technology Council (NSTC), Taiwan, under Project No. 113-2115-M-A49-015-MY2, titled "Investigating Quantum Channels in Noisy Quantum Systems: Some Mathematical Problems in Quantum Information Theory." The authors also acknowledge the assistance of AI-powered language tools in improving the clarity and fluency of the English writing in this manuscript.

\bibliographystyle{ieeetr}
\bibliography{refs}

\end{document}